\documentclass[letter,twocolumn]{jpsj2} 

\def\runauthor{W. Zhang {\it et al.}}

\title{%
High-Field ESR Measurements of S=1/2 Kagome Lattice \\
Antiferromagnet $\mathrm{BaCu_3V_2O_8(OH)_2}$
}

\author{%
Wei-min Zhang$^1$, Hitoshi Ohta$^{1,}$$^2$, Susumu Okubo$^{1,}$$^2$, Masashi Fujisawa$^{2,}$$^3$,  \\
Takahiro Sakurai$^4$, Yoshihiko Okamoto$^5$, Hiroyuki Yoshida$^5$ and Zenji Hiroi$^5$}

\inst{%
$^1$Graduate School of Science, Kobe University, Kobe 657-8501, Japan\\
$^2$Molecular Photoscience Research Center, Kobe University, Kobe 657-8501, Japan\\
$^3$Department of Frontier Research and Technology, Kobe University, Kobe 657-8501, Japan\\
$^4$Center for Supports to Research and Education Activities, Kobe University, Kobe 657-8501,Japan\\
$^5$Institute for Solid State Physics, University of Tokyo, Kashiwa, Chiba 277-8581, Japan\\

}

\recdate{\today}

\abst{%
High-field electron spin resonance (ESR) measurements have been performed on vesignieite $\mathrm{BaCu_3V_2O_8(OH)_2}$, which is considered as a nearly ideal model substance of \textit{S}=1/2 kagome antiferromagnet, in the temperature region from 1.9 to 265 K. The frequency region is from 60 to 360 GHz and the applied pulsed magnetic field is up to 16 T. Observed g-value and linewidth show the increase below 20 K, which suggest the development of the short range order. Moreover, a gapless spin liquid ground state is suggested from the frequency-field relation at 1.9 K.
}

\kword{%
high-field ESR, kagome lattice, frustration, vesignieite, spin dynamics}

\begin{document}
\maketitle
Highly frustrated magnetism has attracted much attention recently. Especially \textit{S}=1/2 kagome lattice antiferromagnet (KAFM) is very important because it is considered to have stronger geometrical frustration than the well studied triangular lattice antiferromagnet~\cite{jps01}. 
Therefore, the ground state of \textit{S}=1/2 KAFM has attracted much interest, but the theoretical studies about its ground state are still controversial. 
The exact diagonalization studies suggest a nonmagnetic spin liquid state with a small spin gap of the order of \textit{J}/20, where \textit{J} is the nearest-neighbor exchange interaction~\cite{jps02}. 
On the other hand, the other numerical calculation studies suggest the gapless ground state~\cite{jps03}. 
Therefore, the experimental studies using the model substance of \textit{S}=1/2 KAFM are really required, but a few model substances are known at present.

Suggested model substances of \textit{S}=1/2 KAFM at present are $\mathrm{Cu^{2+}}$ minerals, herbertsmithite $\mathrm{ZnCu_3(OH)_6Cl_2}$~\cite{jps04}, volborthite $\mathrm{Cu_3V_2O_7(OH)_2\cdot2H_2O}$~\cite{jps05} and vesignieite $\mathrm{BaCu_3V_2O_8(OH)_2}$~\cite{jps06}. 
Herbertsmithite has a rhombohedral structure with the space group $R\overline{3}m$, and is named "structurally perfect kagome compound"~\cite{jps04} because of the ideal kagome geometry of $\mathrm{Cu^{2+}}$ ions. Although the strong antiferromagnetic exchange interaction is expected in herbertsmithite from the large negative Weiss temperature $\theta$=-300 K~\cite{jps07}, no long range order is observed down to 50 mK from the specific heat measurement~\cite{jps08} suggesting the existence of strong frustration in the system. 
However, the neutron diffraction measurement suggested that about 10\% of the $\mathrm{Cu^{2+}}$ sites in the kagome plane are substituted by $\mathrm{Zn^{2+}}$ ions~\cite{jps09}, which is also supported by the NMR measurement~\cite{jps10}. 
The divergent magnetic susceptibility toward \textit{T}=0 observed in herbertsmithite~\cite{jps04} may be related to the existence of the impurity spins caused by these substitutions. 
On the other hand, no such substitution of $\mathrm{Cu^{2+}}$ sites in the kagome plane is expected in the model substance volborthite because a mutual exchange between $\mathrm{Cu^{2+}}$ and $\mathrm{V^{5+}}$ ions is unfavorable in terms of the ionic radius and the Madelung energy. 
The existence of the strong frustration is suggested because the magnetic susceptibility shows no long range order down to 60 mK, but exhibits a tiny spin glass transition at 1.1 K in spite of the large negative Weiss temperature $\theta$=-115 K~\cite{jps11,jps12
}.
The magnetic susceptibility shows a broad maximum at 21 K and it shows a gapless behavior suggested by a large finite value at 60 mK~\cite{jps12}. Very recent magnetization measurement using the high quality polycrystalline sample revealed the magnetization steps at low temperature~\cite{jps12}. 
However, the volborthite, which has a monoclinic structure with the space group \textit{C}2/m, has a slightly distorted kagome lattice, where the distance between two Cu atoms are 0.303 nm (Cu1-Cu2) and 0.294 nm (Cu2-Cu2) in the corner-shared isosceles Cu1-Cu2-Cu2 triangles~\cite{jps13}. 
Recently Okamoto \textit{et al.} suggested a new model substance vesignieite~\cite{jps06}. Vesignieite is a natural mineral with a monoclinic structure of the space group \textit{C}2/m~\cite{jps14}. 
In vesignieite $\mathrm{Cu^{2+}}$ ions form a nearly perfect kagome lattice and no antisite disorder is expected because it does not contain ions chemically similar to $\mathrm{Cu^{2+}}$. Strong frustration is expected because no long range order is observed down to 2 K in spite of the relatively large negative Weiss temperature $\theta$=-77 K~\cite{jps06}. 
From the fitting of the magnetic susceptibility data at high temperature to the high temperature series expansion calculation for the \textit{S}=1/2 KAFM, the exchange coupling and the g-factor are estimated to be \textit{J}=53 K and g=2.16, respectively. At low temperature the magnetic susceptibility shows a sharp increase which suggests about 7\%  impurity contribution. 
By subtracting the contribution from the impurity, the intrinsic magnetic susceptibility shows a broad maximum around 22 K suggesting the development of the short range order. 
In conclusion, Okamoto \textit{et al}. suggest that vesignieite is a nearly ideal kagome lattice compared to volborthite or herbertsmithite.

On the other hand, high-field electron spin resonance (ESR) measurement provides the information about the spin dynamics of system and the precise measure of spin gap mainly at the $\Gamma$ point. 
Especially the high-field ESR is a powerful means to observe the signal with the linewidth beyond 1 T~\cite{jps15}. However, previous high-field ESR results are different among model substances, which may be reflecting the characteristic features of each substance, that is the substitution of magnetic ion or the deformation of kagome lattice. \textit{S}=3/2 KAFM $\mathrm{SrCr_{x}Ga_{12-x}O_{19}}$ (x=8), which has the substitution of $\mathrm{Cr^{3+}}$ ions with $\mathrm{Ga^{3+}}$ ions in the kagome lattice, shows the g-shift and the linewidth broadening below 80 K~\cite{jps16}. On the other hand, \textit{S}=1/2 KAFM herbertsmithite, which has the substitution of $\mathrm{Cu^{2+}}$ ions with $\mathrm{Zn^{2+}}$ ions, shows almost constant g-value and linewidth down to 1.8 K~\cite{jps17}, while volborthite, which has a deformed kagome lattice, shows a g-shift below 20 K and its frequency-field diagram at 1.8 K shows a small gap of about 40 GHz~\cite{jps18}. In this paper, we will show our high-field ESR results on the nearly ideal \textit{S}=1/2 KAFM vesignieite and discuss about the possible gapless spin liquid state at low temperature.

High-field ESR has been measured using pulsed magnetic fields up to 16 T and in the frequency region from 60 to 315 GHz using Gunn oscillators and multipliers. Observed temperature region is from 1.9 to 265 K. The details of our high field ESR system can be found in Refs.~\cite{jps15,jps19,jps20}. A polycrystalline sample of  $\mathrm{BaCu_3V_2O_8(OH)_2}$ (vesignieite) was prepared by the hydrothermal method. The details of the sample preparation can be found in the reference~\cite{jps06}. The obtained sample is found to be a single phase of vesignieite by the powder X-ray diffraction analysis and the magnetic susceptibility showed the same temperature dependence as reported previously~\cite{jps06}.

\begin{figure}[t]
\begin{center}
\includegraphics[keepaspectratio=true,width=90mm]{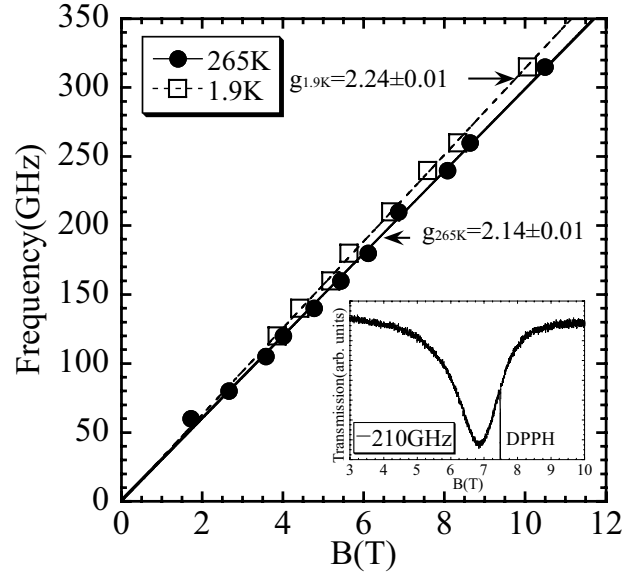}
\end{center}
\caption{Frequency-field relations of ESR observed at 265 K (solid circles) and 1.9 K (open squares). The inset shows the typical ESR spectrum observed at 265 K and 210 GHz where DPPH is the standard of magnetic field.}
\label{f1}
\end{figure}

\begin{figure}[t]
\begin{center}
\includegraphics[keepaspectratio=true,width=90mm]{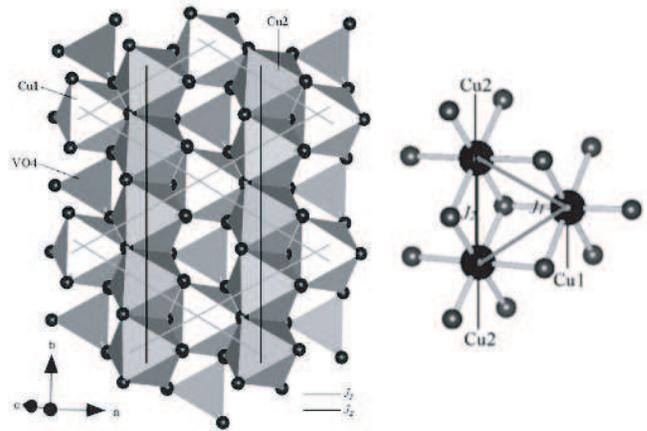}
\end{center}
\caption{Three different $\mathrm{Cu^{2+}}$ sites in the kagome plane of vesignieite.}
\label{f2}
\end{figure}

The inset of Fig.~\ref{f1} shows the typical ESR spectrum observed at 265 K and 210 GHz. Very isotropic ESR absorption line is observed in spite of the fact that the magnetic ions in vesignieite are $\mathrm{Cu^{2+}}$ ions, which typically show an anisotropic powder pattern absorption line reflecting anisotropic g-values~\cite{jps21}. From the crystal structure, there are three different $\mathrm{Cu^{2+}}$ sites in vesignieite as shown in Fig. 2. Each 
$\mathrm{Cu^{2+}}$ site has different principal g-values reflecting each crystal field. 
Therefore, the angular dependence of the ESR absorption line will be different among three $\mathrm{Cu^{2+}}$ sites in Fig.~\ref{f2} if there is no exchange interaction between $\mathrm{Cu^{2+}}$ sites. However, as there exist strong antiferromagnetic exchange interactions between $\mathrm{Cu^{2+}}$ sites, the ESR absorption line will be averaged out by the exchange narrowing effect~\cite{jps22}.  In order to obtain the average g-value of vesignieite at 265 K, frequency dependence measurement is performed. The resonance field of the ESR absorption line is plotted in Fig. 1 and the obtained g-value is $2.14{\pm}$0.01, which is consistent with the g-value of 2.16 obtained from the analysis of the magnetic susceptibility by Okamoto \textit{et al}.~\cite{jps06}. 

\begin{figure}[t]
\begin{center}
\includegraphics[keepaspectratio=true,width=70mm]{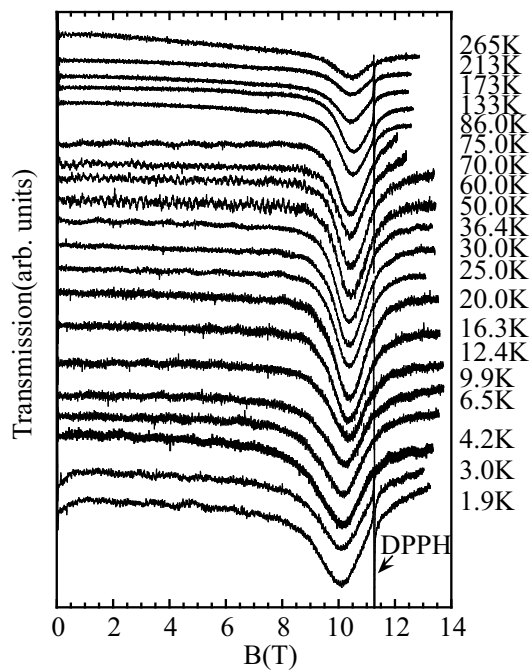}
\end{center}
\caption{Temperature dependence of ESR spectra observed at 315 GHz. DPPH is the standard of magnetic field.}
\label{f3}
\end{figure}

Fig.~\ref{f3} shows the temperature dependence of ESR spectra observed at 315 GHz. As we used different cryostats and detectors above 86 K and below 86 K~\cite{jps19, jps20}, the signal to noise ratio is different. It is clear that the resonance field shifts to lower field and the broadening of absorption line occurs at low temperature. Similar results are also obtained at 160 GHz. 
Temperature dependences of g-values and linewidths obtained at 160 and 315 GHz are summarized in Fig. 4 and Fig.~\ref{f5}, respectively. Here we should point out that the linewidth in Fig. 5 is the linewidth of powder pattern lineshape and  includes also the distribution of g-value because we are dealing with the polycrystalline sample. However, these results suggest that observed ESR is intrinsic and not coming from the paramagnetic impurity observed in the low temperature magnetic susceptibility because no g-shift and no broadening of linewidth at low temperature is expected for the paramagnetic impurity. We think that the contribution of paramagnetic impurity is buried under the broad feature of intrinsic ESR. 

\begin{figure}[t]
\begin{center}
\includegraphics[keepaspectratio=true,width=70mm]{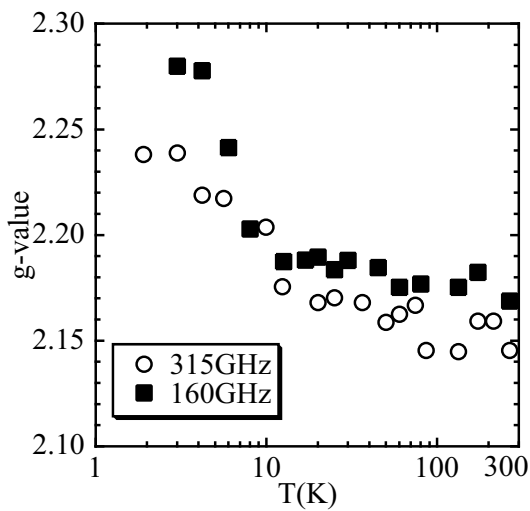}
\end{center}
\caption{Temperature dependence of g-values observed at 160 and 315 GHz.}
\label{f4}
\end{figure}

\begin{figure}[t]
\begin{center}
\includegraphics[keepaspectratio=true,width=70mm]{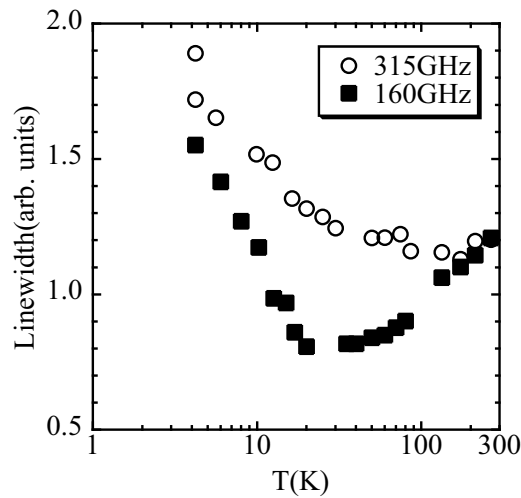}
\end{center}
\caption{Temperature dependence of linewidths observed at 160 and 315 GHz.}
\label{f5}
\end{figure}

Fig.~\ref{f4} shows that the temperature dependences of g-values are similar for all frequencies, which suggest no sign of gap opening at low temperature, and they start to increase below around 20 K. This temperature is closely related to the temperature where the intrinsic magnetic susceptibility shows a broad maximum. Therefore, the g-shifts below around 20 K can be considered as the result of the development of short range order~\cite{jps23} in vesignieite. On the other hand, the temperature dependences of the linewidths seem to show the frequency dependence. For instance, the linewidth observed at 160 GHz decreases as the temperature decreases and starts to increase below around 30 K while the linewidth observed at 315 GHz is almost constant at high temperature and it starts to increase below 30 K. The overall feature resembles the temperature dependence of linewidth in the case of  one dimensional antiferromagnet discussed by Ajiro \textit{et al}.~\cite{jps24}. From the standard ESR theory~\cite{jps24-2}, the linewidth is related to the spin correlation function, and the linewidth is expected to increase as the spin correlation length increases irrespective of the dimensionality. Therefore, the increase of linewidth below 30 K may be also related to the development of short range order in the system. However, as the linewidth in Fig. 5 is the linewidth of powder pattern lineshape, the discussion remains as a future problem.

\begin{figure}[t]
\begin{center}
\includegraphics[keepaspectratio=true,width=70mm]{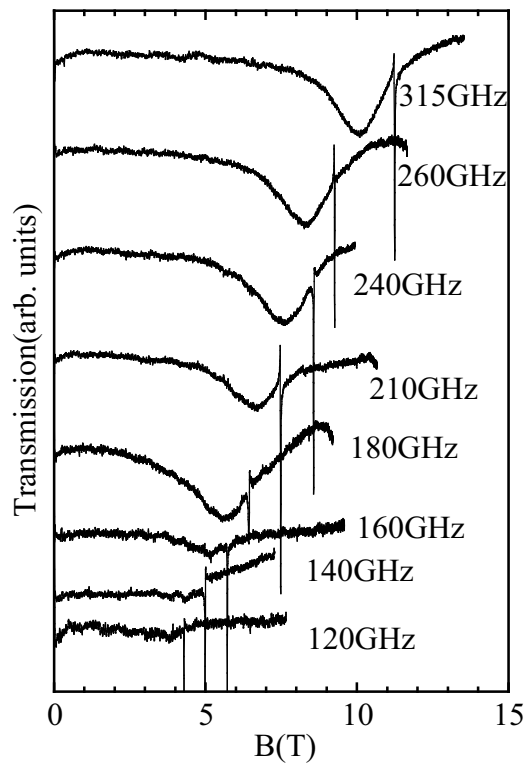}
\end{center}
\caption{Frequency dependence of ESR spectra observed at 1.9 K.}
\label{f6}
\end{figure}

Fig.~\ref{f6} shows the frequency dependence of ESR spectra observed at 1.9 K. As the linewidth increases at low temperature as shown in Fig. 5, the observation of ESR absorption line becomes difficult below 120 GHz. Frequency-field diagram at 1.9 K is plotted in Fig. 1 and the result suggests that vesignieite is still in the paramagnetic state at 1.9 K because it can be interpreted by the linear dependence which crosses the origin. Obtained g-value is $2.24{\pm}$0.01, which is still in a reasonable value range for the paramagnetic resonance of $\mathrm{Cu^{2+}}$ ion. The result strongly suggests that the ground state at 1.9 K is in the spin liquid like state without a gap. However, we cannot exclude the possibility of a spin gap of \textit{J}/20=2.7 K (56 GHz) suggested from the exact diagonalization study~\cite{jps02} because we are not able to observe ESR below 120 GHz (5.4 K).

Recently spin liquid like ground state in the spin frustrated system is suggested experimentally in \textit{S}=1/2 triangular lattice antiferromagnet $\kappa$-$\mathrm{(BEDT-TTF)_2Cu_2(CN)_3}$ by NMR~\cite{jps25}. Although the N\'{e}el order is suggested from the exact diagonalization study of \textit{S}=1/2 Heisenberg triangular lattice system with nearest-neighbor antiferromagnetic exchange interaction~\cite{jps26}, it is considered that the spin liquid like state observed in $\kappa$-$\mathrm{(BEDT-TTF)_2Cu_2(CN)_3}$ may be related to the Mott-insulator physics of the system~\cite{jps27}. Therefore, an investigation of spin liquid ground state in the geometrically frustrated band insulator, not a Mott-insulator, is an another issue. Moreover, the spin gap exists or not in the spin liquid has been a long issue since the proposal of resonating valence bond (RVB) in the triangular lattice antiferromagnet by Anderson~\cite{jps28}. Spin liquid ground state is expected not only in triangular lattice antiferromagnet but also in \textit{S}=1/2 KAFM because the long range order is prohibited from its strong frustration. Although the existence of a spin gap in \textit{S}=1/2 KAFM is suggested by several theoretical calculation~\cite{jps02, jps29}, the temperature dependences of magnetic susceptibility of herbertsmithite~\cite{jps04}, volborthite~\cite{jps05} and vesignieite~\cite{jps06} do not support the existence of spin gap. In case of herbertsmithite, there is a discussion that the spin gap is zero because the system is in the quantum critical point D$_c$=0.1\textit{J}~\cite{jps30} where D is the magnitude of D vector of Dzyaloshinsky-Moriya interaction determined by high frequency ESR~\cite{jps31}. However, it is difficult to believe that volborthite and vesignieite are accidentally at the same quantum critical point with zero spin gap. On the other hand, dynamical properties of herbertsmithite~\cite{jps17} and volborthite~\cite{jps18} observed from ESR show different behaviors. In case of herbertsmithite it shows almost constant $g$-value and linewidth down to 1.8 K~\cite{jps17}. They may be due to the substitution of $\mathrm{Cu^{2+}}$ ions with $\mathrm{Zn^{2+}}$ ions in the kagome lattice but the interpretation remains as a future issue because the $g$-shift and the linewidth broadening are observed in \textit{S}=3/2 KAFM $\mathrm{SrCr_{x}Ga_{12-x}O_{19}}$ (x=8) which has a similar substitution of  $\mathrm{Cr^{3+}}$ ions with $\mathrm{Ga^{3+}}$ ions in the kagome lattice~\cite{jps16}. On the other hand, volborthite is a very clean polycrystalline system~\cite{jps12} and it is a good model substance in that sense. Its ESR shows g-shift below 20 K with frequency dependence~\cite{jps18}. The frequency dependence comes from the formation of small gap of about 40 GHz at 1.8 K. Whether the formation of gap mode is related to the small deformation in the kagome lattice is still under consideration. Therefore, as vesignieite is a nearly ideal \textit{S}=1/2 KAFM, we suggest that ESR behaviors observed in vesignieite is the intrinsic nature of \textit{S}=1/2 KAFM. 

In summary, high-field ESR measurements have been performed on nearly ideal \textit{S}=1/2 KAFM $\mathrm{BaCu_3V_2O_8(OH)_2}$ (vesignieite). High field ESR result in the temperature region from 1.9 to 265 K suggests the development of the short range order below 20 K. Frequency-field relation at 1.9 K strongly suggested that the ground state without a gap at 1.9 K, and the possibility of spin liquid ground state is discussed.

\section{Acknowledgments}

The authors would like to acknowledge Prof. H. Kawamura, Prof. M. Oshikawa and Prof. T. Sakai for valuable discussions. This work was partly supported by a Grant-in-Aid for Scientific Research (C) No. 19540367 from the Japan Society for the Promotion of Science (JSPS) and Grants-in-Aid for Scientific Research on Priority Areas (No. 19052005 "Novel States of Matter Induced by Frustration", No. 17072005 "High Field Spin Science in 100T") from the Ministry of Education, Culture, Sports, Science and Technology (MEXT) of Japan.

\end{document}